\documentclass[12pt]{article}
\usepackage{graphics}
\usepackage{amssymb}
\usepackage{amsmath}

\newcommand{\rf}[1]{(\ref{#1})}
\newcommand{\beq}{\begin{equation}}
\newcommand{\eeq}{\end{equation}}
\newcommand{\bea}{\begin{eqnarray}}
\newcommand{\eea}{\end{eqnarray}}

\newcommand{\e}{\mbox{e}}
\renewcommand{\d}{\mbox{d}}
\newcommand{\g}{\gamma}
\newcommand{\G}{\Gamma}

\renewcommand{\b}{\beta}
\renewcommand{\a}{\alpha}

\newcommand{\m}{\mu}

\newcommand{\ep}{\varepsilon}

\newcommand{\sg}{\sigma}

\newcommand{\oh}{\frac{1}{2}}

\newcommand{\ra}{\rangle}
\newcommand{\la}{\langle}

\newcommand{\mi}{\!-\!}
\newcommand{\equ}{\!=\!}
\newcommand{\pl}{\!+\!}

\newcommand{\cT}{{\cal T}}

\newcommand{\cO}{{\cal O}}

\newcommand{\cB}{{\cal B}}

\newcommand{\hm}{{\hat{m}}}
\newcommand{\hn}{{\hat{n}}}

\newcommand{\no}{\nonumber}

\newcommand{\cm}{{\rm cardy}}
\newcommand{\zz}{{\rm zz}}
\newcommand{\fzz}{{\rm fzz}}
\newcommand{\rara}{\ra\!\ra}
\newcommand{\vv}{|}

\begin{document}

\begin{center}
\vspace{24pt} { \large \bf The nature of ZZ branes}

\vspace{30pt}

{\sl J. Ambj\o rn}$\,^{a,c}$
and
{\sl J. A. Gesser}$\,^{a}$.

\vspace{24pt}
{\footnotesize

$^a$~The Niels Bohr Institute, Copenhagen University\\
Blegdamsvej 17, DK-2100 Copenhagen \O , Denmark.\\
{ email: ambjorn@nbi.dk, gesser@nbi.dk}\\

\vspace{10pt}

$^c$~Institute for Theoretical Physics, Utrecht University, \\
Leuvenlaan 4, NL-3584 CE Utrecht, The Netherlands.\\

\vspace{10pt}

}
\vspace{48pt}

\end{center}


\begin{center}
{\bf Abstract}
\end{center}

In minimal non-critical string theory we show that the principal
(r,s) ZZ brane can be viewed as the basic (1,1) ZZ boundary state
tensored with the (r,s) Cardy boundary state. In this sense there
exists only one ZZ boundary state, the basic (1,1) boundary state.

\vspace{12pt}
\noindent


\newpage

\section{Introduction}

Two-dimensional Euclidean AdS$_{2}$ (the pseudosphere) was quantized
using Liouville quantum field theory by Zamolodchikov and
Zamolodchikov. In \cite{zz} they generalized the quantization of
Liouville theory on the disk \cite{fzz} to the non-compact geometry
of the pseudosphere. The main difference between the quantization of
the disk and the pseudosphere is the assumption with regard to the
pseudosphere that the two-point correlation function factorizes when
the geodesic distance separating the two operators diverges.

Using conformal bootstrap methods the Zamolodchikovs found a number
of conformal invariant boundary conditions, that may be imposed at
``infinity'' of the pseudosphere, and that are consistent with the
above assumption. These boundary conditions were labeled by two
positive integers $(\hm,\hn)$, where the ``basic'' $(1,1)$ boundary
condition played a role quite similar to the $(1,1)$ Cardy boundary
state in the minimal conformal field theories.

In the context of $(p,q)$ minimal non-critical string theory the
boundary conditions of the Zamolodchikovs were given an
interpretation as branes, the so-called ZZ branes
\cite{martinec,sei1,sei2}. In \cite{sei1} it was shown that any ZZ
brane in $(p,q)$ minimal string theory can be viewed as a linear
combination of $(p\mi 1)(q\mi 1)/2$ ``fundamental'' ZZ branes, the
so-called principal ZZ-branes . While the articles
\cite{martinec,sei1,sei2} provided a lot of new understanding of the
target space structure of the ZZ branes and \cite{aagk,ag2} of the
world sheet aspects of ZZ branes, the question raised by the
Zamolodchikovs in their original article remains unanswered. They
said: ``The most intriguing point is the nature of the "excited"
vacua.....A meaning of these quantum excitations of the (physically
infinite faraway) absolute remains to be comprehended''.

We will show that in $(p,q)$ minimal non-critical string theory the
principal ZZ brane labeled by $(\hm,\hn)$ 
has the simple interpretation as the basic $(1,1)$ ZZ boundary state
tensored with the $(\hm,\hn)$ Cardy matter boundary state\footnote{The 
possibility of a relation like \rf{1} was first noticed in \cite{sei2} when
they calculated the ZZ-FZZT amplitude.}:
\beq\label{1} |1,1\ra_{\cm} \otimes |\hm,\hn\ra_{\zz} =
|\hm,\hn\ra_{\cm} \otimes |1,1\ra_{\zz}. \eeq Eq.\ \rf{1} should be
understood in the following way: With regard to expectation values
of physical observables it does not matter whether we use the right
hand side or the left hand side of eq.\ \rf{1}. Thus, there exists
only one ZZ boundary state, the basic $(1,1)$ boundary state, the
other principle ZZ branes being matter-dressed $(1,1)$ boundary
states. Furthermore, we will provide evidence for the following
generalization of \rf{1}: \beq\label{2} |k,l\ra_{\cm} \otimes
|\hm,\hn\ra_{\zz} = \left(\sideset{}{'}\sum_{i=|k-\hm|+1}^{{\rm
top}(k,\hm;p)} \sideset{}{'}\sum_{j=|l-\hn|+1}^{{\rm top}(l,\hn;q)}
|i,j\ra_{\cm}\right) \otimes |1,1\ra_{\zz}, \eeq where
\bea\label{2a}
{\rm top}(k,\hm;p)&=&\min(k+\hm-1,2p-1-k-\hm), \\
{\rm top}(l,\hn;q)&=&\min(l+\hn-1,2q-1-l-\hn),\label{2b} \eea
Notice, this summation\footnote{The prime in the summation symbol
$\sideset{}{'}\sum $ means that the summation runs in steps of two.}
is precisely the same which appears in the fusion of two primary
operators in the $(p,q)$ minimal conformal field theory:
\beq\label{3} O_{k,l} \times O_{\hm,\hn} =
\sideset{}{'}\sum_{i=|k-\hm|+1}^{{\rm top}(k,\hm;p)}
\sideset{}{'}\sum_{j=|l-\hn|+1}^{{\rm top}(l,\hn;q)}  [O_{i,j}].
\eeq

Why are eqs.\ \rf{1} and \rf{2} true?

Recall the definition of the Cardy matter boundary states in the
$(p,q)$ minimal conformal field theory: \beq\label{3a} |k,l\ra_\cm
\equiv \sum_{i,j} \frac{S(k,l;i,j)}{\sqrt{S(1,1;i,j)}}\; \vv
i,j\rara \eeq
where the summation runs over all the different Ishibashi states
$\vv i,j\rara$ in the $(p,q)$ minimal model, \beq\label{3b}
S(k,l;i,j) = 2 \sqrt{\frac{2}{pq}} (-1)^{1+kj+li}\sin (\pi b^2
lj)\;\sin(\pi k i /b^2) \eeq is the modular S-matrix in the $(p,q)$
minimal model and $b^2\equ p/q$.
The Cardy matter boundary states are labeled by two
integers $(k,l)$, which satisfy that $1 \leq k \leq p\mi 1$, $1\leq
l \leq q\mi 1$ and $ kq-lp > 0$.

On the other hand the principal ZZ boundary states are defined as
\beq\label{3c} |\hm ,\hn\ra_{\zz} = \int_0^\infty \d P \;
\frac{\sinh ( 2\pi \hm P/b) \; \sinh (2\pi \hn P b)}{\sinh ( 2\pi
P/b) \; \sinh ( 2\pi P b)}  \; \Psi_{1,1} (P) \, \vv P \rara, \eeq
where 
$b \equ \sqrt{p/q}$. $\Psi_{1,1}(P)$ is the basic $(1,1)$ ZZ wave
function \cite{zz}: \beq\label{3d} \Psi_{1,1}(P) = \b \frac{iP
\m^{-iP/b}}{\G(1-2iPb)\G(1-2iP/b)}, \eeq where the constant $\b$ is
independent of the cosmological constant $\m$ and $P$. Finally, $\vv
P\rara$ denotes the Ishibashi state corresponding to the non-local
primary operator
$\exp(2(Q/2+iP)\phi)$ in Liouville theory, where $Q\equ b \pl 1/b$.
The principal ZZ branes are labeled by two integers $(\hm,\hn)$,
where $1 \leq \hm \leq p\mi 1$, $1\leq \hn \leq q\mi 1$, $\hm q \mi
\hn p
> 0$.

Notice, the ranges of the indices $k,l$ labeling the different Cardy
matter boundary states and the indices $\hm,\hn$ labeling the
principal ZZ branes are the same. As noted already by the
Zamolodchikovs in \cite{zz}, the modular bootstrap equations for the
ZZ boundary states are surprisingly similar to the bootstrap
equations for the Cardy matter boundary states in the minimal
models.
The key point is now that the physical operators in minimal string
theory, to be discussed below, carry both a matter ``momentum'' and
a Liouville ``momentum'' and these are not independent, but related
by the requirement that the operators scale in a specific way. In
particular, the Liouville momenta $P$ of the physical observables
are imaginary and the imaginary $i$ explains the shift from $\sin$
to $\sinh$ going from \rf{3a} to \rf{3c}. The coupling between the
matter and Liouville momenta implies, that physical expectation
values will be the same irrespectively of whether we use the left or
the right side of eq.\ \rf{1}.

Below we will verify \rf{1} and \rf{2} for physical bulk operators
evaluated on ${\rm AdS}_2$,  as well as \rf{2} with regard to the
FZZT-ZZ cylinder amplitude and the ZZ-ZZ cylinder amplitude.

\section{The disk amplitude}

According to \cite{sei1} the physical operators in minimal
non-critical string theory are the tachyon operators, the ground ring
operators and the copies of the ground ring operators at negative ghost
number.

The tachyon operators are given by \beq\label{4} \cT_{r,s} =
c\bar{c}\; \cO_{r,s} \; \e^{2\b_{r,s} \phi}, \eeq where $c$ is the
ghost field, $\cO_{r,s}$ the primary matter operators, \beq\label{5}
\b_{r,s} = \oh \left( Q -r/b+s\,b\right),~~~~~rq-ps > 0, \eeq and
\beq\label{6} Q=b+1/b,~~~~~b=\sqrt{p/q}. \eeq

In order to provide evidence for \rf{1} we calculate the tachyon
one-point function on the pseudosphere with the states $|k,l\ra_\cm
\otimes |1,1\ra_\zz $ and $|1,1\ra_\cm \otimes |k,l\ra_\zz $ imposed
at infinity.

Since the Liouville ``momentum'' corresponding to $\cT_{r,s}$ is
$P_{r,s} = i(Q/2-\b_{r,s})$ one obtains from \rf{3c}: \beq\label{7}
\la \cT_{r,s}| \Big( |k,l\ra_\cm \otimes |1,1\ra_\zz \Big) = \a
\frac{S(k,l;r,s)}{\sqrt{S(1,1;r,s)}} \; \Psi_{1,1}(P_{r,s}), \eeq
and \beq\label{8} \la \cT_{r,s}| \Big( |1,1\ra_\cm \otimes
|k,l\ra_\zz \Big) = \a  \sqrt{S(1,1;r,s)} \; \frac{\sinh(2\pi k
P_{r,s}/b) \sinh(2\pi l P_{r,s} b)}{ \sinh(2\pi  P_{r,s}/b)
\sinh(2\pi P_{r,s} b)} \Psi_{1,1}(P_{r,s}), \eeq where $\a$
is a constant independent of $r,s,k$ and $l$.

Using the expression  \rf{3b} for the modular S-matrix it is now
simple algebra to verify that the right hand sides of eqs. \rf{7}
and \rf{8} are equal.

One can also arrive at this conclusion starting from a FZZT boundary
state tensored with a Cardy matter state, $|k,l\ra_\cm \otimes
|\sg\ra_\fzz$, where $\sg$ is related to the boundary cosmological
constant $\m_b$ and the bulk cosmological constant $\m$ by
\beq\label{9a} \frac{\m_b}{\sqrt{\m}}= \cosh (\pi b \sg). \eeq
According to \cite{sei1} (by a calculation similar to the one
leading to \rf{7} and \rf{8}) one has: \beq\label{9} \la \cT_{r,s}
|(|k,l\ra_\cm \otimes |\sg\ra_\fzz) = A_{r,s} (-1)^{ks+lr}
\cosh\left(\frac{\pi\sg(r-sb^2)}{b}\right) \sin\left(\frac{\pi
kr}{b^2}\right)\sin \left(\pi ls b^2\right) \eeq Using
\cite{martinec} \beq\label{10} |k,l\ra_\zz = |i(k/b+lb)\ra_\fzz -
|i(k/b-lb)\ra_{\fzz}, \eeq which is valid for principal ZZ boundary
states, one can now calculate both $\la \cT_{r,s}| ( |k,l\ra_\cm
\otimes |1,1\ra_\zz)$ and $\la \cT_{r,s}| ( |1,1\ra_\cm \otimes
|k,l\ra_\zz)$ and verify that they agree. For future reference we
present one of the calculations: \bea
\lefteqn{\la \cT_{r,s}| ( |k,l\ra_\cm \otimes |1,1\ra_\zz)}\no\\
&=&
\la \cT_{r,s}| ( |k,l\ra_\cm \otimes (|i(1/b\pl b)\ra_\fzz-
|i(1/b-b)\ra_\fzz ))\label{10a}\\
&=&-2 A_{r,s} (-1)^{ks+lr}\sin\left(\frac{\pi
kr}{b^2}\right)\sin(\pi ls b^2)
\sin\left(\frac{\pi r}{b^2}\mi\pi s\right)\sin(\pi r \mi\pi sb^2) \label{10b}\\
&=&2 A_{r,s} (-1)^{ks+lr-s-r}\sin\left(\frac{\pi kr}{b^2}\right)\;
\sin(\pi ls b^2) \; \sin\left(\frac{\pi r}{b^2}\right)\; \sin(\pi
sb^2) \label{10c} \eea A similar calculation of $\la \cT_{r,s}| (
|1,1\ra_\cm \otimes |k,l\ra_\zz)$ gives the same result.

The ground ring operators as well as the physical operators at
negative ghost number have FZZT one-point functions which are
similar to the tachyon one-point functions \rf{9}, except that the
range of $s$ is different from the one for
the tachyon operators \cite{sei1}. Using calculations similar to
\rf{10a}-\rf{10c} one can thus verify that the expectation values of
these operators are independent of which of the two branes in eq.\
\rf{1}
we impose at infinity.

We finally turn to the verification of \rf{2}. Let $\cB_{r,s}$
denote a physical operator, i.e.\ a tachyon operator, a ground ring
operator or one of the copies of a ground ring operator at negative
ghost number. In this case $s<q$, $s \neq 0$ mod $q$ and $qr-ps>0$.
These operators satisfy \rf{9} with $\cB$ substituted for $\cT$. Our
task is to calculate the matrix element \beq\label{11} \la
\cB_{r,s}|(|k,l\ra_\cm \otimes |\hm ,\hn\ra_\zz). \eeq Using \rf{10}
for $ |\hm ,\hn\ra_\zz$ and then \rf{9} and the following identity
\beq\label{11a} \sinh (nx) \sinh (n'x) = \sum_{l=0}^{\min(n,n')-1}
\sinh (x) \sinh ((n+n'-2l-1)x) \eeq one obtains after some algebra:
\beq\label{12} \la \cB_{r,s}|(|k,l\ra_\cm \otimes |\hm ,\hn\ra_\zz)
= \sideset{}{'}\sum_{i=|k-\hm|+1}^{k+\hm-1}
\sideset{}{'}\sum_{j=|l-\hn|+1}^{l+\hn-1} \la \cB_{r,s}|(|i,j\ra_\cm
\otimes |1,1\ra_\zz) \eeq This is still not in agreement with \rf{2}
since the range of summation does not always agree with \rf{2}.
However, assume that $\hm\pl k > p$. We now split the sum over $i$
in two at $i=2p \mi 1 \mi\hm\mi k$ in accordance to \rf{2}.
With regard to the second part of the sum we then obtain using
\rf{10b}: \beq\label{13} \sideset{}{'}\sum_{i=2p+1-\hm-k}^{k+\hm-1}
\sideset{}{'}\sum_{j=|l-\hn|+1}^{l+\hn-1} \la \cB_{r,s}|(|i,j\ra_\cm
\otimes |1,1\ra_\zz) \propto
\sideset{}{'}\sum_{i=2p+1-\hm-k}^{k+\hm-1} (-1)^{is}\sin(\pi\,
ir/b^2) =0 \eeq Similar arguments for $\hn\pl l > q$ leads to
\beq\label{14} \sideset{}{'}\sum_{i=|\hm-k|+1}^{k+\hm-1}
\sideset{}{'}\sum_{j=2q+1-\hn-l}^{l+\hn-1} \la
\cB_{r,s}|(|i,j\ra_\cm \otimes |1,1\ra_\zz) \propto
\sideset{}{'}\sum_{j=2q+1-\hn-l}^{l+\hn-1} (-1)^{jr}\sin(\pi\,
js\,b^2) =0 \eeq Hence, the range of summation in \rf{12} may be
expressed as in \rf{2}.
\section{The cylinder amplitude}

To simplify the discussion we restrict ourselves to $(2,2m\mi 1)$
minimal string theory. However, we expect the result to be valid in
any $(p,q)$ minimal string theory. In $(2,2m\mi 1 )$ non-critical
string theory the Cardy matter boundary states are labeled by only
one integer $r$, which satisfies $1 \leq r \leq m-1$. Our basic
object is the FZZT-FZZT cylinder amplitude $Z(r,\sg';s,\sg)$
obtained in \cite{ag2}.
The principal $(\hm,\hn)$ ZZ boundary states in the $(2,2m\mi 1)$
minimal string theories have $\hm \equ 1$ and $1 \leq \hn \leq m-1$.
We may calculate the FZZT-ZZ cylinder amplitude from \rf{10}:
\beq\label{15a} Z(r,(1,\hn)_\zz;s,\sg) = Z(r,\sg'\equ i(1/b \pl\hn
b);s,\sg) -Z(r,\sg' \equ i(1/b\mi \hn b);s,\sg). \eeq Let us now
define the following quantity:
\bea\label{15}
\lefteqn{K(r,\sg';s,\sg) \equiv}\\
&& \sqrt{2} \pi^2 \int_{\g_u} \d P \; \frac{P}{P^2+\ep^2} \;
\frac{e^{2\pi i \sg P} \cos(2\pi \sg' P) \;\sinh(2\pi rb P)
\;\sinh(2\pi sbP)}{ \sinh(4\pi P/b) \;\sinh^2(2\pi bP)} \no \\
&+& \sqrt{2} \pi^2 \int_{\g_l} \d P \;\frac{P}{P^2+\ep^2} \; \frac{
e^{-2\pi i \sg P} \cos(2\pi \sg' P) \;\sinh(2\pi rb P) \;\sinh(2\pi
sbP)}{ \sinh(4\pi P/b) \;\sinh^2(2\pi bP)}, \no \eea where $\g_u$
is a path enclosing the singularities in the upper half plane
anti-clockwise, while $\g_l$
is a path enclosing the singularities in the lower half plane
clockwise. We have followed \cite{sei2} and used the regularization
$$
\frac{1}{P} \to \frac{P}{P^2+\ep^2}
$$
in \rf{15}.
The FZZT-FZZT cylinder amplitude is actually given by \rf{15} for
$r+s \leq m$. For $0 < \sg' < \sg$ this may be seen by deforming the
integration contour appearing in the integral expression for the
FZZT-FZZT cylinder amplitude in \cite{ag2}. However, the integral
expression in \cite{ag2} is not convergent for generic values of
$\sg$ and $\sg'$ and we therefore define the cylinder amplitude for
generic values of $\sg$ and $\sg'$ by the analytic continuation of
\rf{15}.\footnote{If one performs the summation over residues, one
realizes that \rf{15} is actually symmetric in $\sg$ and $\sg'$.}
Hence, one has \beq\label{16} K(r,\sg';s,\sg) =
Z(r,\sg';s,\sg)~~~{\rm for}~~~ r+s \leq m. \eeq As argued in
\cite{ag2}, for $r+s
> m$ the expression for $Z(r,\sg';s,\sg)$ is more complicated as a
simple consequence of the fusion rules \rf{3} in the matter sector.
This complication does not affect the FZZT-ZZ cylinder amplitude in
the sense that the additional terms present in $Z(r,\sg';s,\sg)$
compared to $K(r,\sg';s,\sg)$ for $r+s>m$ cancel in the difference
\rf{15a}. We thus have \beq\label{17} Z(r,(1,\hn)_\zz;s,\sg) =
K(r,\sg'\equ i(1/b\pl\hn b);s,\sg) -K(r,\sg'\equ i(1/b\mi \hn
b);s,\sg) \eeq for all values of the Cardy matter labels $r$ and
$s$.

We can now use \rf{17} and \rf{15} to perform a calculation similar
to the ones already performed in \rf{10a}-\rf{10c} and
\rf{11}-\rf{12}. The algebraic operations are performed on the
$K$-integrands and using \rf{11a} it is easily seen that
\bea\label{19}
\lefteqn{Z(r,(1,\hn)_\zz;s,\sg)}\hspace{2cm}\\
&=& \sideset{}{'}\sum_{k=|r-\hn|+1}^{r+\hn-1} K(k,\sg'\equ i(1/b\pl
 b);s,\sg) -K(k,\sg'\equ i(1/b\mi b);s,\sg)
\nonumber\\
&=& \sideset{}{'}\sum_{k=|r-\hn|+1}^{r+\hn-1} Z(k,(1,1)_\zz;s,\sg).
\nonumber \eea which is consistent with \rf{2}.

We finally consider the ZZ-ZZ cylinder amplitude. Using \rf{10} and
\rf{19} one finds: \beq\label{20} Z(r,(1,\hn)_\zz;s,(1,\hm)_\zz)=
\sideset{}{'}\sum_{k=|r-\hn|+1}^{r+\hn-1}
\sideset{}{'}\sum_{l=|s-\hm|+1}^{s+\hm-1}
Z(k,(1,1)_\zz;l,(1,1)_\zz), \eeq again in accordance with \rf{2}.

\section{Discussion}

We have provided some evidence that \rf{1} and \rf{2} are valid for
physical expectation values in minimal non-critical string theory.
It answers partly the question raised by the Zamolodchikovs and
quoted in the introduction. In the cases where one actually has a
concrete model like the $(p,q)$ minimal conformal field theory as
the matter content in non-critical string theory, one can view the
principal ZZ branes as matter-dressed basic (1,1) ZZ boundary
states.
Because of the special form \rf{3b} of the fusion matrix, imposing a
Cardy matter condition at infinity different from the basic $(1,1)$
Cardy matter condition is equivalent to (at least for the
observables we have considered) multiplying the basic ZZ wave
function $\Psi_{1,1}(P)$ with a factor: \beq\label{21} \Psi_{1,1}(P)
\to \frac{\sinh ( 2\pi \hm P/b) \; \sinh (2\pi \hn P b)}{\sinh (
2\pi P/b) \; \sinh ( 2\pi P b)} \;\Psi_{1,1}(P) \equiv
\Psi_{\hm,\hn}(P), \eeq $\Psi_{\hm,\hn}(P)$ being the wave function
of the principal $(\hm,\hn)$ ZZ boundary state. The Zamolodchikovs
derived the set of wave functions $\Psi_{\hm,\hn}(P)$ by demanding
that the two-point function factorize into a product of one-point
functions when the geodesic distance diverges, but they only
considered Liouville theory, i.e.\ the matter part had been
integrated out. What we have shown in this article is that the
effect of imposing a $(\hm,\hn)$ Cardy matter boundary condition at
infinity is precisely to produce a $\Psi_{\hm,\hn}(P)$ wave function
in Liouville theory. The integral expression for the FZZT-ZZ
cylinder amplitude in \rf{17} and \rf{15} is indeed obtained after
integrating over the matter degrees of freedom.

Within the context of Liouville theory it was observed in \cite{zz}
that the two-point functions diverge as a function of the geodesic
distance on the pseudosphere if one consider a $(\hm,\hn)$ ZZ brane
with $(\hm,\hn) \neq (1,1)$. It is interesting if this can be
understood in terms of Cardy matter boundary states. In principle we
have the machinary to address such questions not only in terms of
the ``background geometry'' of the pseudosphere, but in terms of the
full quantum geometry using the so-called loop-loop propagator
$G_\m(l_1,l_2;d)$. \cite{kawai1,kawai2,gk,aw_pq}. As shown in
\cite{ag2} it allows us to describe the transition from compact to
non-compact geometry as the geodesic distance $d \to\infty$, and in
some cases it also allows us to relate very explicitly to the matter
states at the boundaries \cite{kawai2,akw,aajk}. In the so-called
CDT-model of 2d quantum gravity \cite{al}, which is closely related
to ordinary Euclidean 2d quantum gravity \cite{ackl}, we have
already observed a drastic change in the behavior of the correlation
functions due to the geometry of the pseudosphere \cite{ajwz} and
the same thing might happen in the full Euclidean theory.

\section*{Acknowledgment}
Both authors acknowledge the  support by
ENRAGE (European Network on
Random Geometry), a Marie Curie Research Training Network in the
European Community's Sixth Framework Programme, network contract
MRTN-CT-2004-005616.

\end{document}